\documentclass[12pt]{article}
\usepackage{graphicx}
\def\PRL{Phys. Rev. Lett. }
\def\PRD{Phys. Rev. D }
\def\JCAP{J. Cosmol. Astropart. Phys. }
\def\nuall{\nu^{^{\hbox{\hspace*{-3mm}{\tiny (---)}}}}}

\newcommand\pubnumber{NuPhys2015-Gil-Botella}
\newcommand\pubdate{\today}

\def\ciemat{CIEMAT, Department of Basic Research,\\
Av. Complutense, 40, 28040 Madrid, SPAIN}

\def\Title#1{\begin{center} {\Large #1 } \end{center}}
\def\Author#1{\begin{center}{ \sc #1} \end{center}}
\def\Address#1{\begin{center}{ \it #1} \end{center}}

\newcommand\pubblock{\rightline{\begin{tabular}{l} \pubnumber\\
         \pubdate  \end{tabular}}}
\newenvironment{Abstract}{\begin{quotation}  }{\end{quotation}}
\newenvironment{Presented}{\begin{quotation} \begin{center} 
             PRESENTED AT\end{center}\bigskip 
      \begin{center}\begin{large}}{\end{large}\end{center} \end{quotation}}

\def\beq{\begin{equation}}
\def\eeq#1{\label{#1}\end{equation}}
\def\eeqn{\end{equation}}

\def\beqa{\begin{eqnarray}}
\def\eeqa#1{\label{#1}\end{eqnarray}}
\def\eeqan{\end{eqnarray}}

\let\bar=\overbar

\def\Dslash{\not{\hbox{\kern-4pt $D$}}}
\def\dslash{\not{\hbox{\kern-2pt $\del$}}}

\def\msb{{\bar{\ssstyle M \kern -1pt S}}}

\begin{document}
\begin{titlepage}
\pubblock

\vfill
\Title{Detection of Supernova Neutrinos}
\vfill
\Author{In\'es Gil-Botella}
\Address{\ciemat}
\vfill
\begin{Abstract}
The neutrino burst from a core-collapse supernova can provide information about the star explosion mechanism and the mechanisms of proto neutron star cooling but also about the intrinsic properties of the neutrino such as flavor oscillations. One important question is to understand to which extent can the supernova and the neutrino physics be decoupled in the observation of a single supernova. The capabilities of present and future large underground neutrino detectors to yield information about the time and flavor dependent neutrino signal from a future galactic supernova are described in this paper. Neutrinos from past cosmic supernovae are also observable and their detection will improve our knowledge of the core-collapse rates and average neutrino emission. A comparison between the different experimental techniques is included.
\end{Abstract}
\vfill
\begin{Presented}
NuPhys2015, Prospects in Neutrino Physics\\
Barbican Centre, London, UK,  December 16--18, 2015
\end{Presented}
\vfill
\end{titlepage}
\def\thefootnote{\fnsymbol{footnote}}
\setcounter{footnote}{0}
%

\section{Introduction}

Supernova (SN) neutrinos are part of the main physics program for future large underground neutrino projects. Flavor composition, energy spectrum and time structure of the neutrino burst can provide information about the SN properties (core-collapse, explosion, neutron star cooling mechanism) \cite{mirizzi}. Despite the enormous recent progress on SN simulations \cite{tamborra}, much about the physics of core collapse is not well understood. Therefore, the detection of neutrinos from the next galactic burst will provide critical information needed to improve our knowledge about the collapse dynamics.

The neutrino signal from a galactic SN can also give information about the intrinsic properties of neutrinos such as flavor oscillations, in particular, the mass hierarchy \cite{mirizzi}. In addition, because neutrinos arrive before any other signal from a SN, it is possible to provide an early alert to the astronomical community combining the observations of several experiments \cite{snews}. This will allow an early observation of the first stages of the SN explosion.

The detection of the Diffuse Supernova Neutrino Background (DSNB) from all past core-collapse SN in the Universe will be a major discovery and will help to understand the stellar population and provide an independent test of the SN rate \cite{lunardini}.

\section{The Supernova Neutrino Signal}

Core-collapse SN are a huge source of neutrinos of all flavors in our Universe. During a SN explosion, 99\% of the gravitational binding energy of the star (of the order of 3 x 10$^{53}$ ergs)  is released by neutrinos and antineutrinos of all flavors, which play the role of astrophysical messengers, escaping from the SN core.

Neutrinos from core-collapse supernovae are emitted in a burst of a few tens of seconds duration \cite{mirizzi}. Essentially three stages can be distinguished: the neutronization burst, a large $\nu_e$ emission in the first 10's of ms, the accretion phase, which lasts between few tens to few hundreds of ms, where the luminosity of electron and non-electron flavors is significantly different, and the cooling phase, up to $\sim$10 s, when luminosities and average energies decrease and there is almost luminosity equipartition between neutrino species.

\begin{figure}[htbp]
\centering
\includegraphics[width=0.55\linewidth]{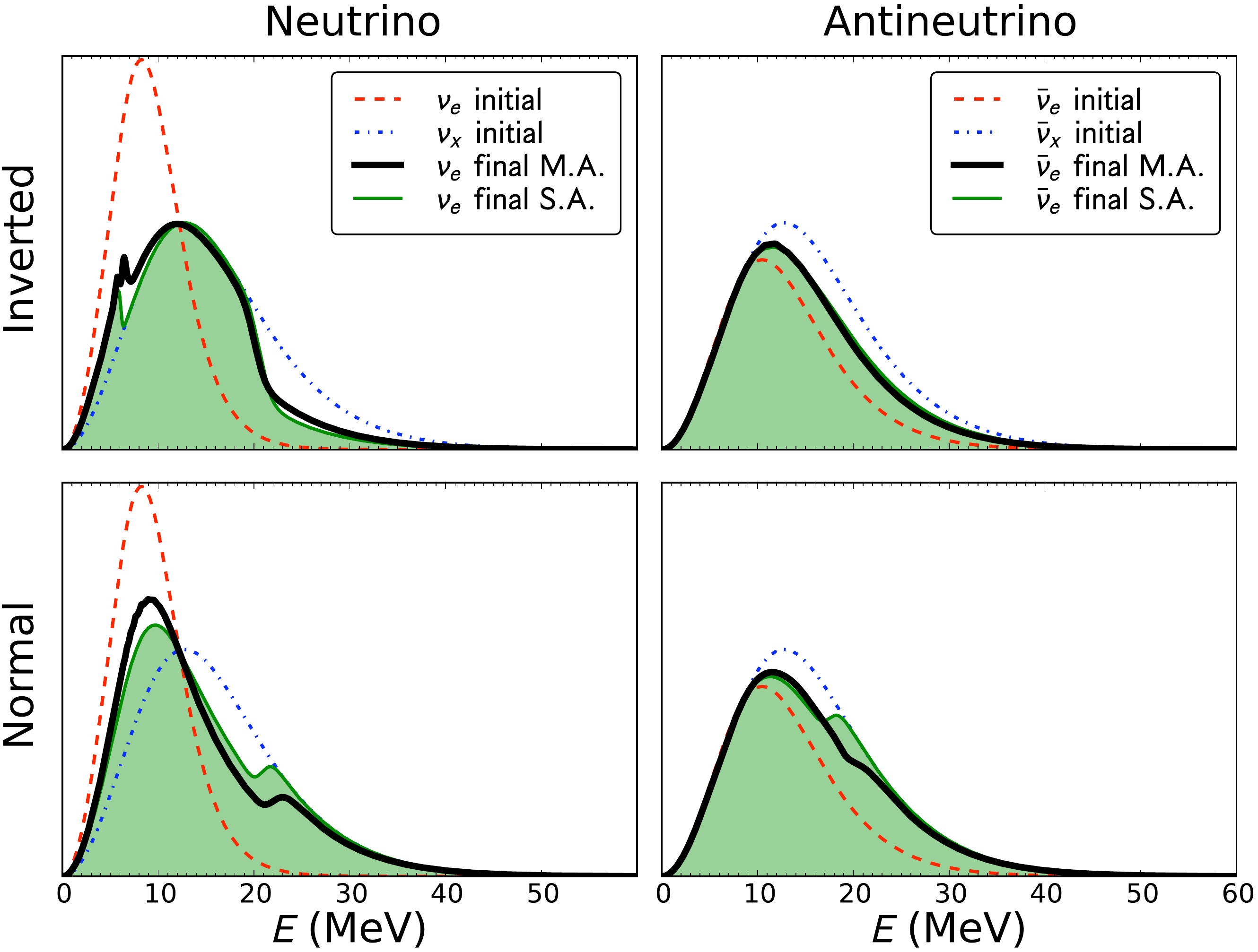}
\caption{Spectra of $\nu_e$ (left) and $\bar\nu_e$ (right), for both inverted IH (top) and normal NH (bottom) hierarchies. Multiangle (M.A.) results are shown with thick solid curves, single-angle (S.A.) with filled regions and initial spectra with dashed lines (from \cite{friedland}).}
\label{fig:friedland}
\end{figure}

An example of the expected supernova neutrino spectra integrated over 10 s can be seen in Fig. \ref{fig:friedland} for the different flavor components. However, MSW flavor transitions and other neutrino collective effects may modify significantly these spectra, as shown in Fig. \ref{fig:friedland}. By looking at the particular time and spectra features, information about the SN mechanisms and neutrino properties can be obtained.

The basic core-collapse model was confirmed in 1987 by the observation of neutrino events from the SN1987A~\cite{sn1987}, which exploded in the Large Magellanic Cloud at a distance of $\sim$50 kpc. This was the first and so far the only time when stellar collapse neutrinos were observed. The burst of light was visible to the naked eye. Around three hours before the observation of the SN, an increase of neutrinos was detected by two water ($\sim$1 kt) Cherenkov detectors: Kamiokande, IMB and one liquid scintillator detector: Baksan. Around 20 events were detected in a 13 seconds interval at a time consistent with the estimated time of the core collapse. Within uncertainties, all signals are contemporaneous. The detection of this SN served to confirm the predictions about light curves and production of neutrinos. However, limited quantitative information on the neutrino spectrum was obtained due to the small statistics recorded.

The combination of Kamiokande and IMB data allowed to define confidence regions in the binding energy and spectral $\bar\nu_e$ temperature plane~\cite{raffelt}. This observation allowed us to put strong constraints on exotic neutrino properties and limits on non-standard cooling mechanisms associated with new particles emitted from the SN core (right-handed neutrinos and axions). However, with the poor statistics, determining both SN and neutrino parameters was very difficult. 

\section{Experimental requirements to detect supernova neutrinos}

The next generation of large underground neutrino detectors will be designed to allow the detection of core-collapse SN neutrinos. The SN neutrino detectors must have a large mass and a low energy threshold of the order of a few MeVs. Ideally, they must be sensitive to all neutrino flavors and to the products of the interactions in the few to few tens of MeV range. In order to measure the time and energy distributions of the different neutrino flavors, detectors must have very good capabilities for neutrino energy resolution, angular and time resolution, and particle identification. 

There are various backgrounds for the SN neutrino detection, and they may vary by the detector location and type as the signal channel. In general, it is not a serious concern for a SN neutrino burst since it lasts only for about 10 s. The most important background sources are natural radioactivity and cosmogenic backgrounds. Other backgrounds are reactor neutrinos, solar neutrinos or low energy atmospheric neutrinos, but they are vey low for SN neutrino bursts.

In order to estimate the neutrino time and energy distributions in our detector, we need to know the expected neutrino flux at the detector from a SN at a certain distance, including MSW and collective flavor oscillations (from the theoretical SN simulations), the cross section of the detection channel, the detector mass (number of targets) and the detector resolution and efficiency.

There exist many different theoretical SN neutrino models that include flavor oscillations with sizeable differences in the predicted neutrino fluxes in terms of energy distributions, time evolution, etc. In addition, the interaction processes of SN neutrinos at low energies ($<$ 100 MeV) are not well known. The most relevant neutrino interactions at these energies are:
\begin{itemize}
\item
Elastic scattering (ES) on electrons ($\nu_x + e^- \to \nu_x + e^-$): the cross section is relatively low compared to the other interactions but it is sensitive to all flavor neutrinos and provides pointing information. The electron is scattered in the direction of the neutrino.
\item
The inverse beta decay (IBD) interaction ($\bar{\nu}_e + p \to  n + e^+$) is the most significant interaction in the current detectors. It has a kinematical threshold of 1.8 MeV. The positron energy loss can be observed and the neutrons are captured producing gammas. This reaction has a high cross section but it is only sensitive to $\bar\nu_e$.
\item
The elastic scattering (ES) on protons ($\nu_x + p \to \nu_x + p$): the total cross section is four times smaller than that of IBD but it sensitive to all flavors. The proton recoil energy is highly suppressed by the nucleon mass so it will be difficult to reconstruct it but within the reach of low-background large scintillation detectors.
\item
Charged-current interactions (CC) with nuclei ($\nu_e + (N, Z) \to (N+1,Z-1) + e^-$ and $\bar{\nu}_e + (N, Z) \to (N+1,Z-1) + e^+$):  these are interactions of $\nu_e$ and $\bar\nu_e$ with neutrons and protons in nuclei. The energy loss of the charged lepton is observable. Nucleons and gammas produced by the final nucleus as it deexcites may also be observable and may help to tag the interaction.
\item
Neutral-current interactions (NC) with nuclei ($\nu_x + A \to \nu_x + A^*$): these interactions produce observable signals via ejected nucleons or de-excitation gammas. The coherent elastic scattering on nuclei produce recoil energies in the few keV range out of reach of conventional detectors (dark matter detectors should be sensitive).
\end{itemize}

Unfortunately, only a few of the neutrino interactions relevant for current detectors in the tens-of-MeV range have precisely known cross sections. Except for ES and IBD, the theoretical and experimental cross section knowledge is very limited. 

\section{Supernova neutrino detectors: present and future}

There are essentially three detector technologies to measure SN neutrinos: Liquid Scintillator (LSc), Water Cherenkov (WC) and Liquid Argon (LAr) TPC detectors. Three large volume underground detectors are being proposed that can guarantee continuous exposure for several decades so that a high statistics supernova neutrino signal could be eventually observed.
In Table \ref{tab:rates} the expected number of events for a 20 kt LSc, 560 kt WC and 40 kt LArTPC detectors are summarized. The technologies are complementary to study the various neutrino flavors and features of the events. LAr TPCs provide fundamental information on the detection of $\nu_e$s not reachable by other technologies, particularly on the initial neutronization burst.

\begin{table}[t]
\begin{center}
\begin{tabular}{lccc}
Detector & Mass (kt) & Events & Main flavor\\  \hline
LSc & 20 & 6000 & $\bar\nu_e$\\
WC & 560 & 110,000 & $\bar\nu_e$ \\
LAr TPC & 40 & 4000 & $\nu_e$\\ \hline
\end{tabular}
\caption{Neutrino event rate estimates for a core-collapse SN at 10 kpc. There may be significant variations by SN model.}
\label{tab:rates}
\end{center}
\end{table}

\subsection{Liquid scintillators}
The main interaction channel for a supernova burst signal in LSc detectors is the IBD, because of the presence of large number of free protons in the liquid. In this process, $\bar\nu_e$s interact with free protons giving a positron and a neutron. Both signals are time and space correlated: prompt signal comes from scintillation and annihilation of the positron and 200 $\mu$s (or 30 $\mu$s) later, the neutron is thermalized and captured in H (or Gd). This delayed coincidence is an excellent signature to tag IBDs. This process has a high cross section and a neutrino energy threshold of 1.8 MeV.

These detectors observe the scintillation light emitted by the interaction of particles in the liquid. They produce a large number of photoelectrons that are collected by PMTs leading to excellent energy resolution and low energy thresholds. The pointing capability is limited but some directional information is possible \cite{lsc_pointing}.

Neutrino interactions on carbon and ES also occur. In particular, the NC excitation of $^{12}$C will produce a 15 MeV de-excitation gamma which may be observable providing a determination of the total neutrino flux. Neutrino-proton elastic scattering can also be observed. This signal produces very low recoil energy protons nevertheless significant amount of events can be observed \cite{beacom}.

The size of the current LSc detectors is $\sim$1 kt so, they can detect around 300 events from a SN at 10 kpc. For the next generation of LSc, like JUNO, RENO-50 or LENA, with masses of the order of 20-50 kt, we could expect around 6000 events from a SN at 10 kpc \cite{JUNO}.

JUNO is also able to constrain the absolute neutrino mass by observing the SN neutrinos \cite{JUNO_mass}. Since the arrival time and neutrino energy can be well measured at JUNO, the distortion in the time distribution of SN neutrino events caused by the delay of flight time is sensitive to the absolute mass scale. Assuming a nearly-degenerate mass spectrum and normal hierarchy, an upper bound is found to be m$_{\nu}$ $<$ 0.83 $\pm$ 0.24 eV at 95\% CL, for a SN at 10 kpc.

\subsection{Water Cherenkov}
The main channel for supernova neutrino detection in WC detectors is the IBD, since the abundance of free protons. Therefore, like LSc detectors, WC detectors are mainly sensitive to $\nu_e$s. They have the advantage that huge detectors are possible. Charged particles are detected via Cherenkov light emission. Neutron tagging is possible through the observation of gamma-Compton scatters, however, due to the Cherenkov threshold, the detection of the 2.2 MeV gamma is difficult.

Since the Cherenkov photons carry directional information, the ES can be used to point to the SN. Other interactions in oxygen nuclei of water contribute to the SN burst signal. De-excitation gammas from NC interactions may be visible but the detection efficiency is very poor.

The largest current WC detector is SK (22.5 kt). The expected SN neutrino rates with a 5 MeV detection energy threshold are 8000 events in total \cite{sk-sn}. It may be possible to dissolve Gd compounds in the water to enhance the neutron tagging: 0.1\% Gd gives $>$ 90\% n-capture efficiency. For the case of the proposed HK experiment\cite{HK-sn}, the SN neutrino rates will increase dramatically up to $\sim$200.000 events for a SN at 10 kpc, being also sensitive to SNs from LMC (50 kpc) with $\sim$10.000 events and even from Andromeda (775 kpc) with $\sim$50 events.

The long-string WC detectors are arrays of PMTs in water or ice (as IceCube or Antares). Although they are nominally designed for the study of very high astrophysical neutrinos, they can also be able to detect SN neutrinos, if the background rates are sufficiently low. Signal will be seen as a coincident increase in the single PMT count rates, above the background counts. This is the case of the IceCube detector which has a SN trigger installed. They cannot reconstruct individual neutrino interaction events, so they are insensitive to spectral and directional information. However, thanks to the large photon statistics, it has very good timing (2 ms), being sensitive to the time structure of the burst. The SN rate at 10 kpc will be $\sim$170,000 events~\cite{icecube}. IceCube could be able to distinguish between NH and IH, depending on the models and the SN distance. For some models and a SN at 10 kpc, a 3$\sigma$ sensitivity to distinguish the mass hierarchy can be obtained.

\subsection{Heavy nuclei detectors}
Neutrino detection based on high Z materials will give the largest possible cross sections and excellent neutrino detection efficiency. Lead has an attractively large neutrino-scattering cross section per nucleon compared with other elements.

The HALO detector at SNOlab \cite{halo} is a dedicated SN detector made by 79 tons of Pb and SNO $^3$He counters. The predominant $\nu_e$CC reaction in HALO produces Bi nuclei in excited states.  It is also sensitive to other neutrino flavors through the neutral current interaction. Both the CC and NC reactions will produce an excited nucleus which will de-excite by emitting one or more neutrons and $\gamma$ rays to reach the ground state. As a result of this similarity between the signals for the CC and NC interactions, the HALO experiment cannot distinguish which interaction has occurred. The released neutrons thermalize by elastic collisions and are detected by the $^3$He proportional counters.

This technique has limitations since no event-by-event energy information or pointing information can be obtained and only rates are provided. For a SN at 10 kpc, $\sim$30 events are expected in the HALO detector. HALO-2 is a proposed upgrade to the kiloton scale.

\subsection{Liquid Argon TPCs}

LAr TPCs have excellent sensitivity to electron neutrinos from SN. This technology provides good energy resolution and full particle reconstruction with very high quality tracking. Energy thresholds as low as a few MeV may be possible. In these detectors, the ionization charge is drifted by an electric field towards the anode where the charge is collected. Using the time arrival of the charge at the readout planes, a three-dimensional track reconstruction is possible. Particles are identified by the rate of energy loss along the track. The Ar scintillation light is also detected enabling fast timing of signals and event localization inside the detector.

At low energy, neutrinos can be detected in LAr through four detection channels.
\begin{enumerate}
\item Charged-current (CC) interactions on argon ($\nu_e+~^{40}Ar \to e^-+~^{40}K^*$ and $\bar{\nu}_e +~^{40}Ar \to e^++~^{40}Cl^*$). The neutrino energy thresholds of these reactions are 1.5 and 7.48 MeV, respectively.
\item Neutral-current (NC) interactions on argon for all flavor neutrinos ($\nuall +~^{40}Ar \to \nuall +~^{40}Ar^*$). The energy threshold of this reaction is 1.46 MeV.
\item Elastic scattering (ES) on atomic electrons for all flavor neutrinos ($\nuall+~e^- \to \nuall+~e^-$)
\end{enumerate}

\begin{figure}[htbp]
\centering
\includegraphics[width=0.5\linewidth]{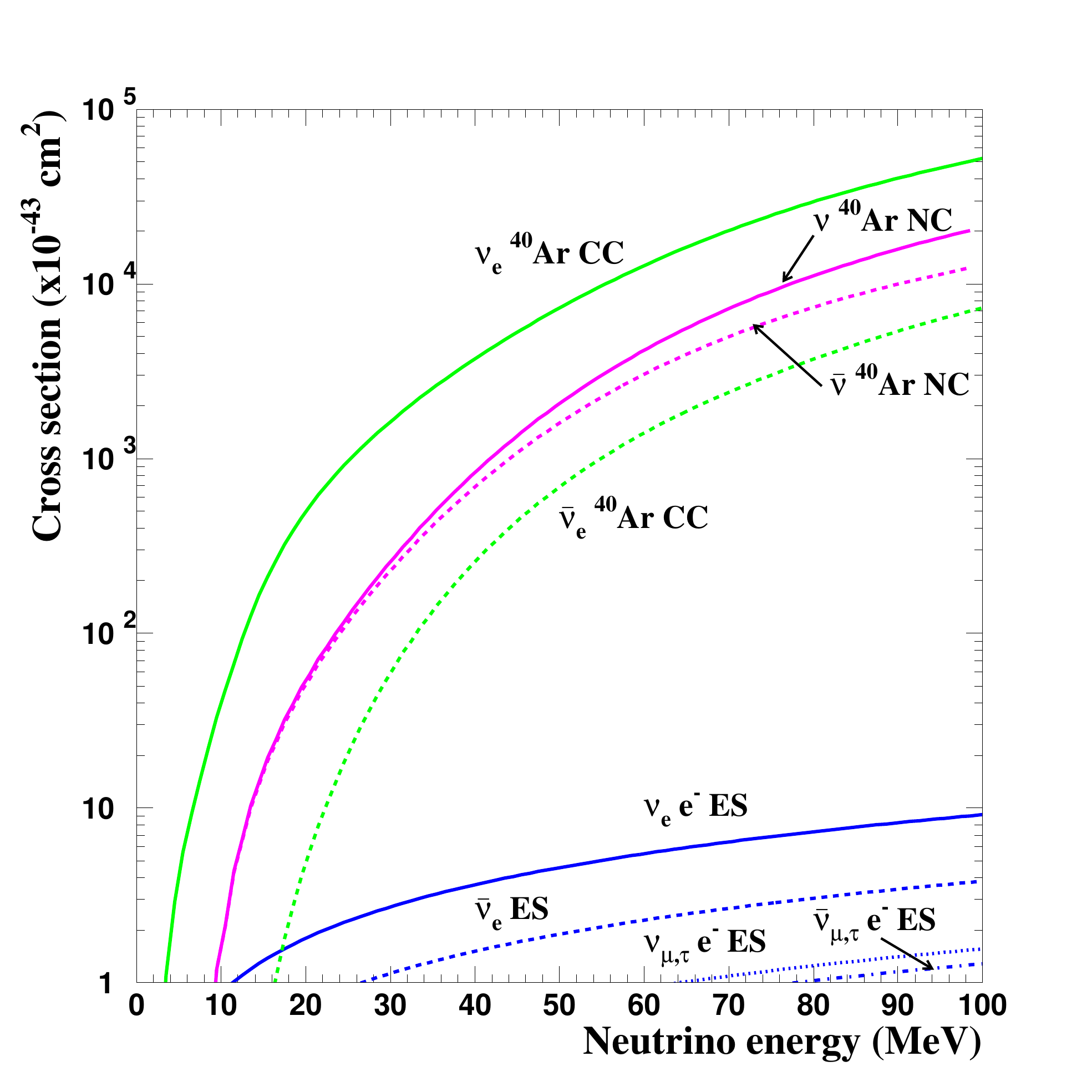}
\caption{Neutrino cross sections relevant to the supernova detection with a liquid argon TPC.} 
\label{fig:cross}
\end{figure}

Figure \ref{fig:cross} shows the cross sections of all the processes as a function of the neutrino energy. These cross sections have been computed by Random Phase Approximation for neutrino energies up to 100 MeV \cite{Gil-Botella:2004bv}. It is possible to separate the different channels by measuring the associated photons coming from the de-excitation of K, Cl and Ar or by the absence of photons in the case of elastic scattering. 

It is important to have a good knowledge of the final state of the interactions, specially the de-excitation $\gamma$s from the $^{40}$K. However, precision models of low energy neutrino argon reactions or dedicated measurements are not available.

The most promising proposal for a large LAr TPC is the DUNE project~\cite{DUNE}. The idea is to build a 40 kt underground LAr TPC detector at 1300 km from Fermilab in the SURF laboratory at $\sim$1500 m depth. Table \ref{tab:DUNE} shows the rates calculated for the dominant interactions in argon for the “Livermore” model \cite{livermore} (no longer preferred, but included for comparison with literature), and the “GKVM” model \cite{Gava:2009pj}; for the former, no oscillations are assumed; the latter assumes collective effects. In general, there is a rather wide variation, up to an order of magnitude, in event rate for different models.

\begin{table}[t]
\begin{center}
\begin{tabular}{lcc}
Channel & Events Liv. model & Events GKVM model \\  \hline
$\nu_e + ^{40}Ar \to e^- + ^{40}K^*$ & 2720 & 3350 \\
 $\bar{\nu}_e +^{40}Ar \to e^+ + ^{40}Cl^*$ & 230 & 160 \\
$\nuall +~e^- \to \nuall+~e^-$ & 350 & 260 \\ \hline
\end{tabular}
\caption{Event rates in DUNE for a core-collapse SN at 10 kpc.}
\label{tab:DUNE}
\end{center}
\end{table}

\begin{figure}[htbp]
\centering
\includegraphics[width=0.43\linewidth]{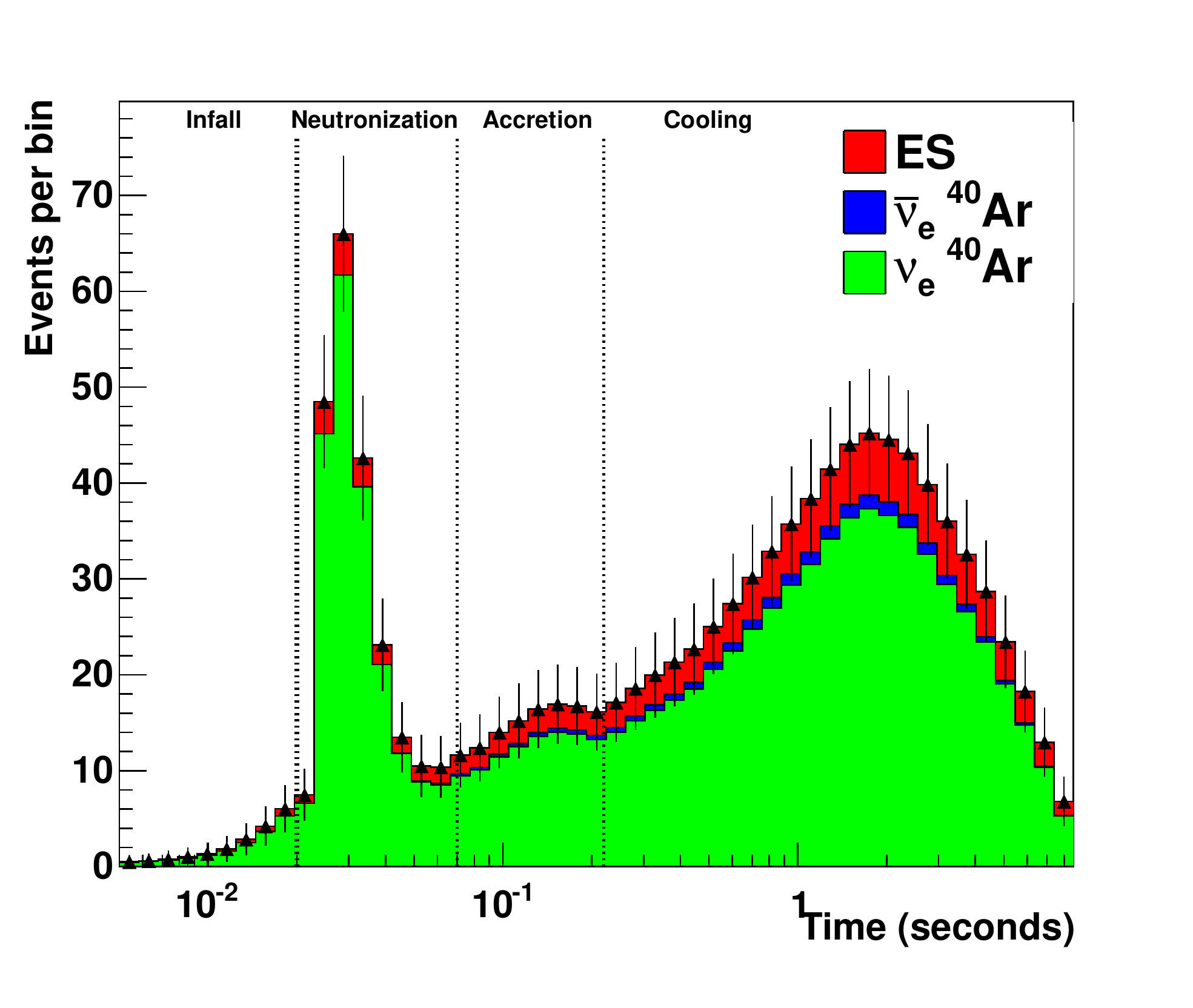}  \hspace{-0.5cm}
\includegraphics[width=0.43\linewidth]{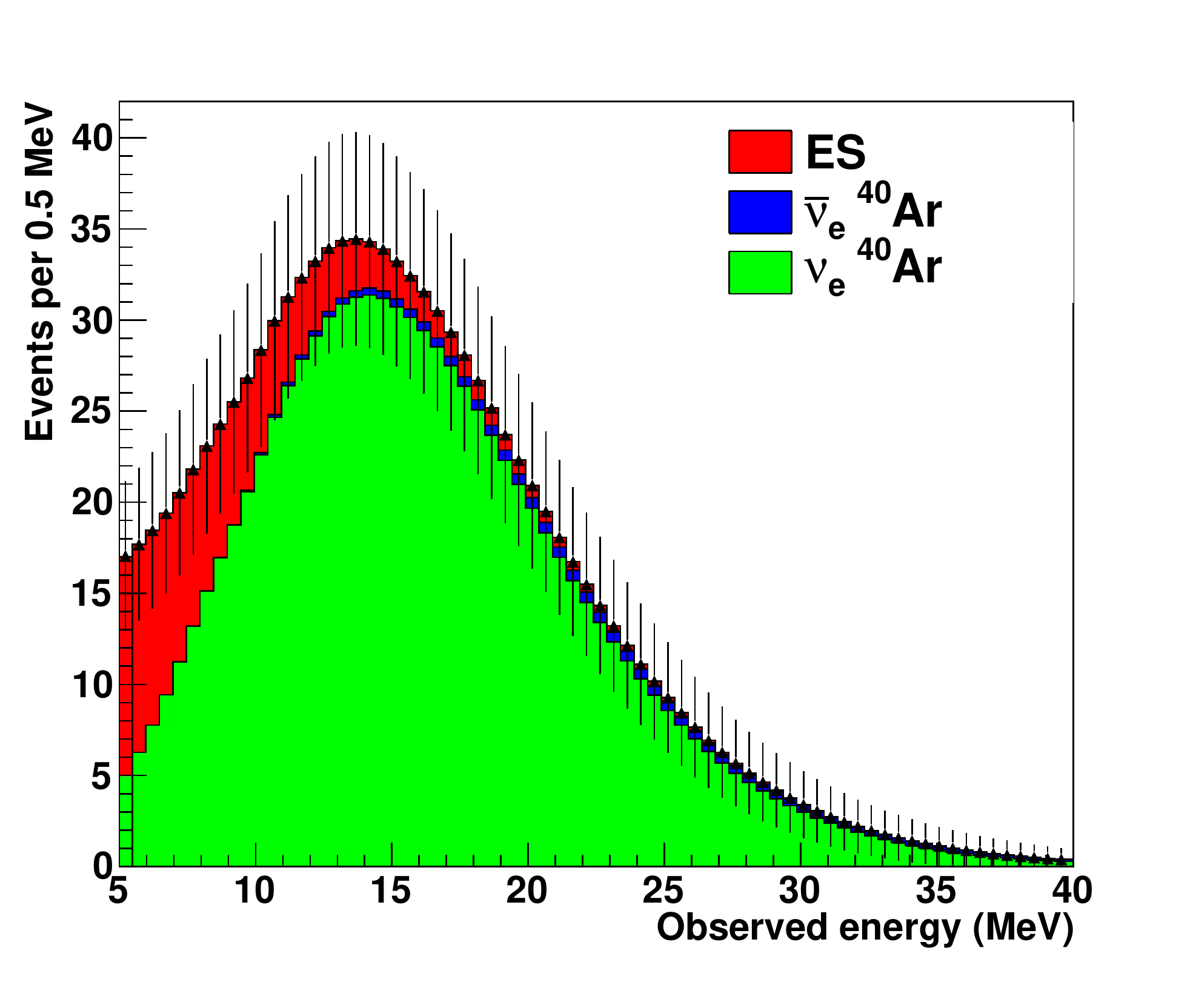}
\caption{Left: Expected time-dependent signal in 40 kt of liquid argon for an electron-capture supernova~\cite{Huedepohl:2009wh} at 10~kpc, calculated using SNOwGLoBES~\cite{snowglobes}, showing breakdown of event channels.  Right: Expected measured event spectrum for the same model, integrated over time.} 
\label{fig:eventrates}
\end{figure}

Fig.~\ref{fig:eventrates} shows the event distribution as a function of time and the expected energy spectrum for the different CC and ES detection channels for a SN at 10 kpc. No oscillations have been considered. We see the clear dominance of the $\nu_e$ channel and the huge visibility of the neutralization burst.

\begin{figure}[htbp]
\centering
\includegraphics[width=0.4\linewidth]{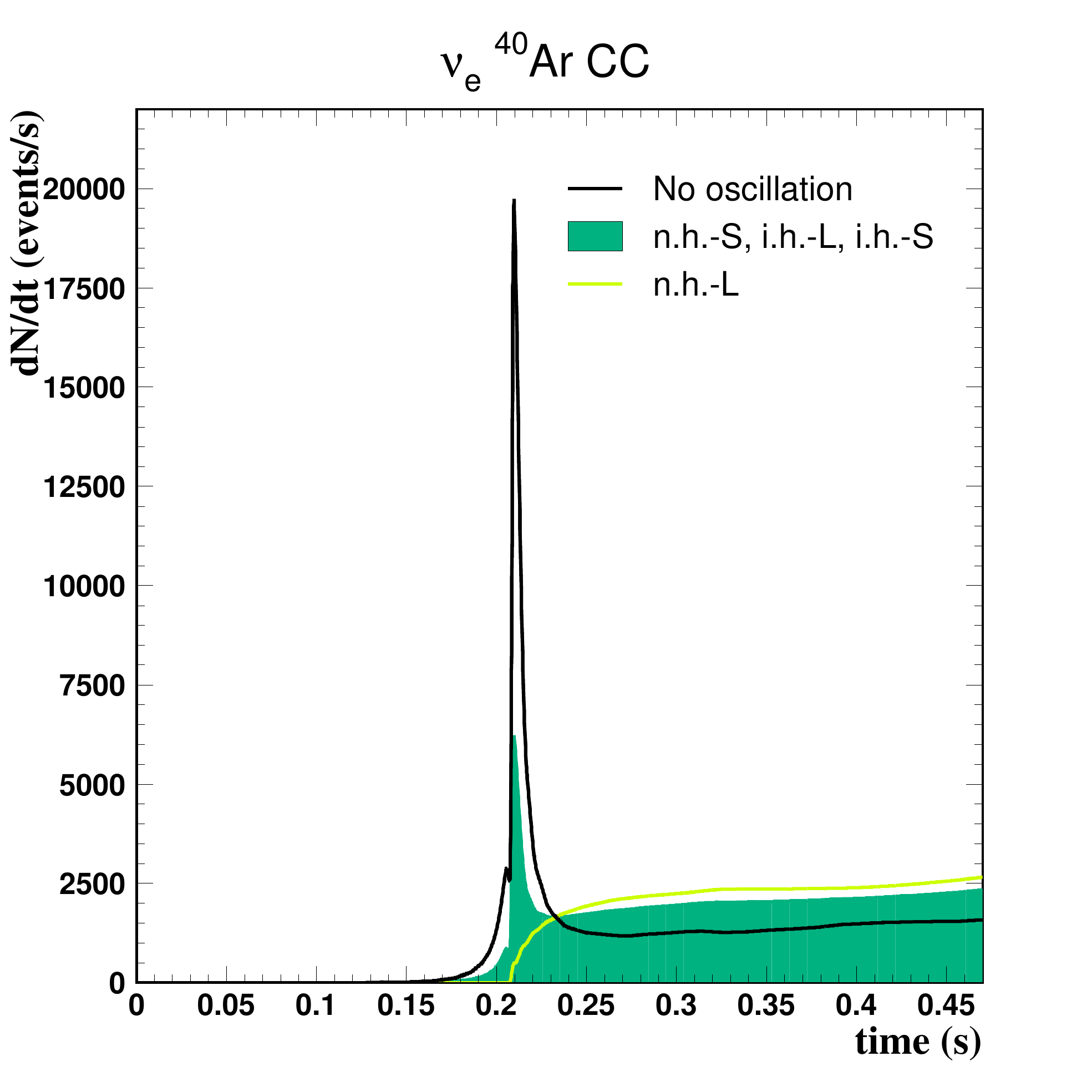} \hspace{-0.5cm}
\includegraphics[width=0.4\linewidth]{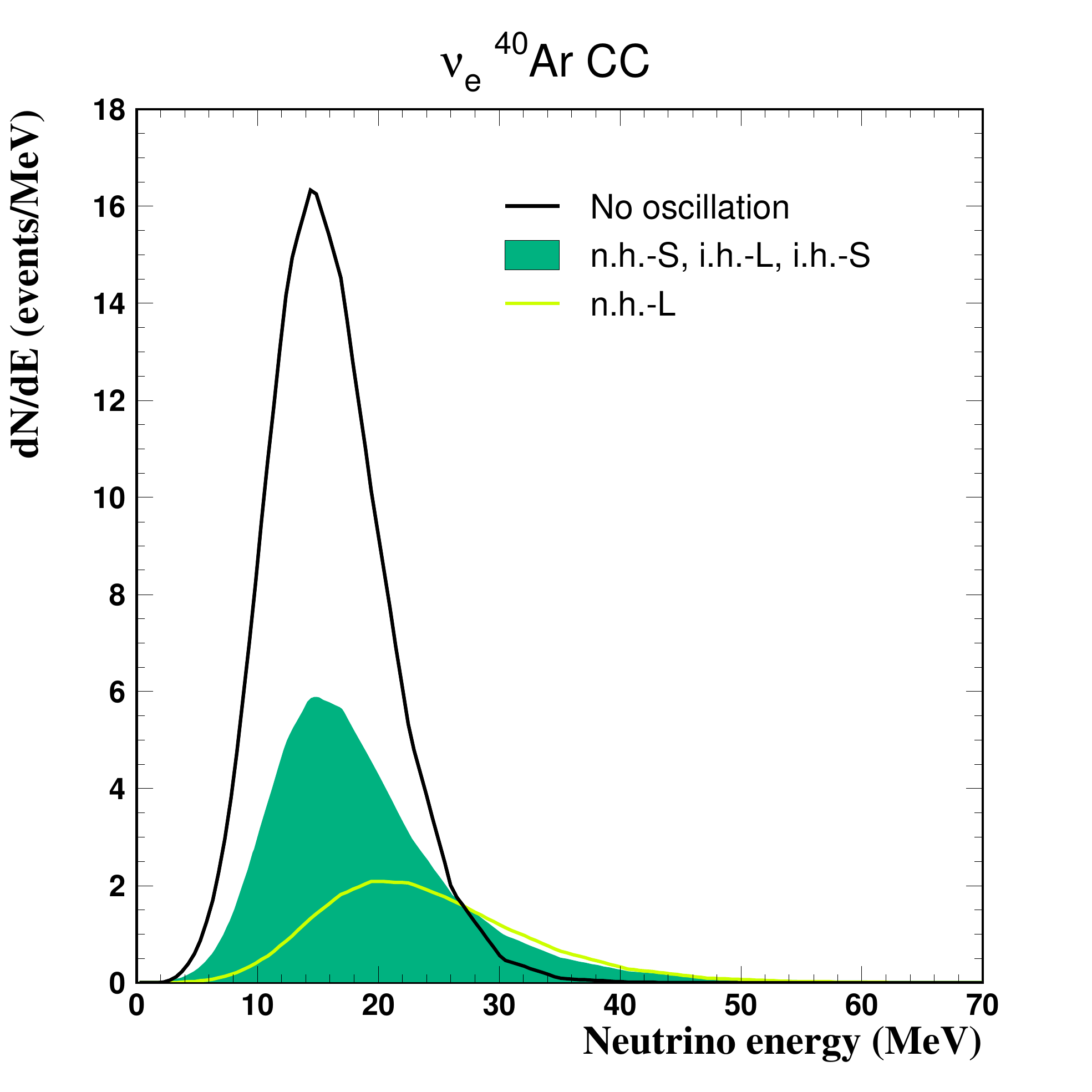}
\caption{Time evolution of the $\nu_e$CC event rate (left) and the corresponding time integrated event spectra (right) for different oscillation scenarios (from \cite{Gil-Botella:2004bv}).} 
\label{fig:early}
\end{figure}

LAr TPCs can provide unique information about the early breakout pulse from the detection of $\nu_e$ neutrinos mainly through the CC process. The advantage of the early phase analysis is that is not affected by the time evolution of the density structure of the star or whether the remnant is a neutron star or a black hole. The analysis of the time structure of the supernova signal during the first few tens of milliseconds after the core bounce can provide a clean indication if the full $\nu_e$ burst is present or absent and therefore allows distinguishing between different mixing scenarios. Fig. \ref{fig:early} shows the effect of the reduction of the $\nu_e$ peak due to oscillations. The suppression is maximal for normal hierarchy and large $\theta_{13}$ mixing angle (n.h.-L) due to the total conversion of $\nu_e$ into $\nu_{\mu,\tau}$. The energy spectrum moves slightly to higher neutrino energy values. From the detector point of view, a good time resolution will be needed.

\begin{figure}[htbp]
\centering
\includegraphics[width=0.5\linewidth, angle=-90]{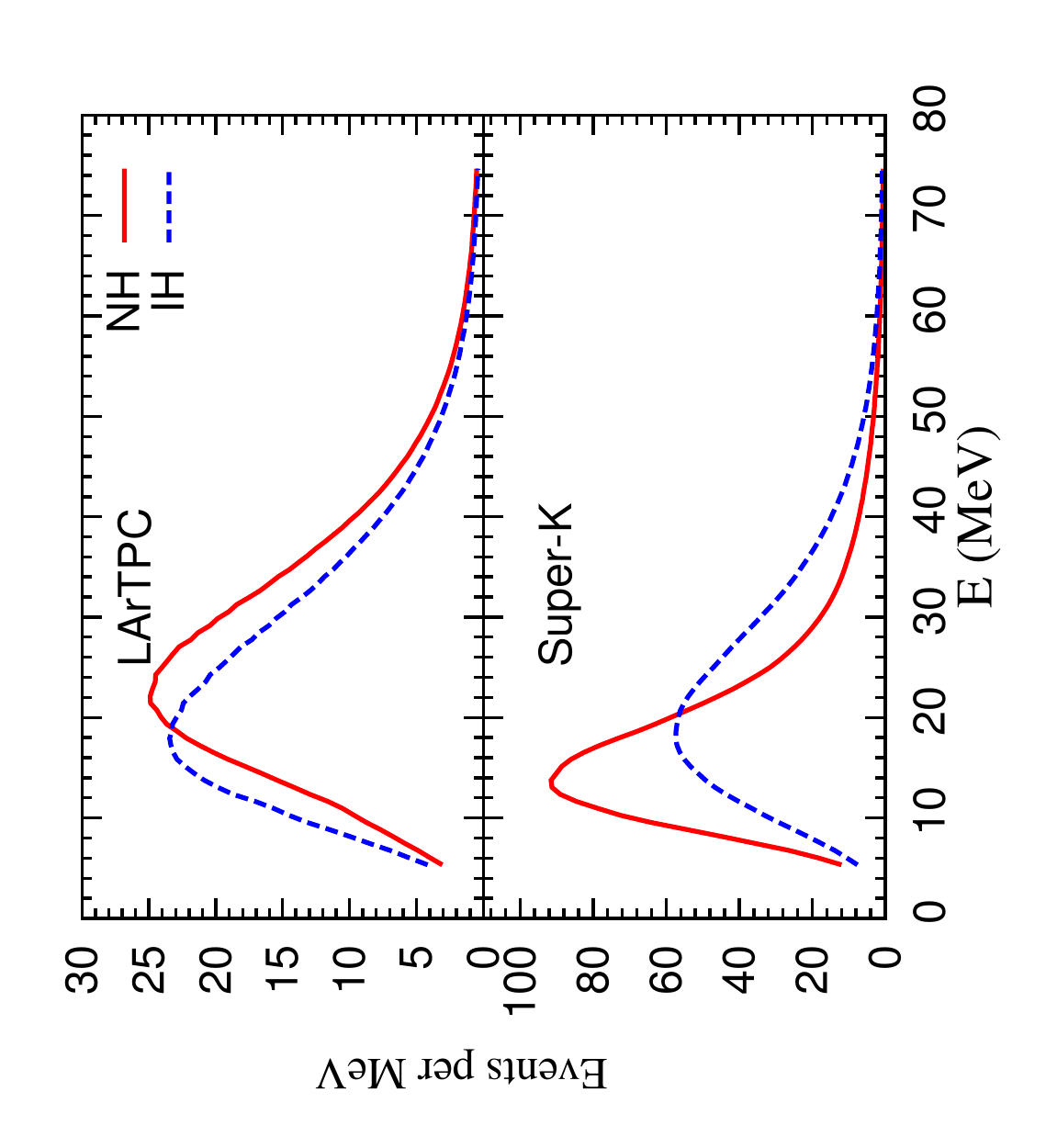}
\caption{Time-integrated energy spectra of neutrino events during the first 0.5~s for a 35 kt LArTPC and the Super-K detector (from \cite{martinez-pinedo}).} 
\label{fig:pinedo}
\end{figure}

The expected neutrino rates as a function of time and the energy spectra of SN neutrinos will be significantly modified by MSW flavor conversions and collective oscillations \cite{Gil-Botella:2004bv}. These observations may help to identify the type of hierarchy. As an example, Fig. \ref{fig:pinedo} shows the time-intergrated spectra during the first 0.5 s for a 35 kt LAr TPC and the SK detector. The NH could be identified if a large difference in the neutrino spectra is detected between both detectors.

\section{Diffuse Supernova Neutrino Background}

The diffuse supernova neutrino background (DSNB) is the flux of neutrinos and antineutrinos emitted by all core-collapse supernovae occurred so far in the Universe. It will appear isotropic and time independent in feasible observations. The detection of the DSNB will be an important discovery and will provide valuable information on the astrophysics of core collapse, failed SNe and black-hole formation, and flavor-dependent neutrino propagation and flavor conversion.

For all detection technologies, backgrounds are the main limiting factors, as they often restrict the sensitive energy window considerably and even in the energy window they limit the benefits of the larger detector mass.

\begin{figure}[htbp]
\centering
\includegraphics[width=0.55\linewidth]{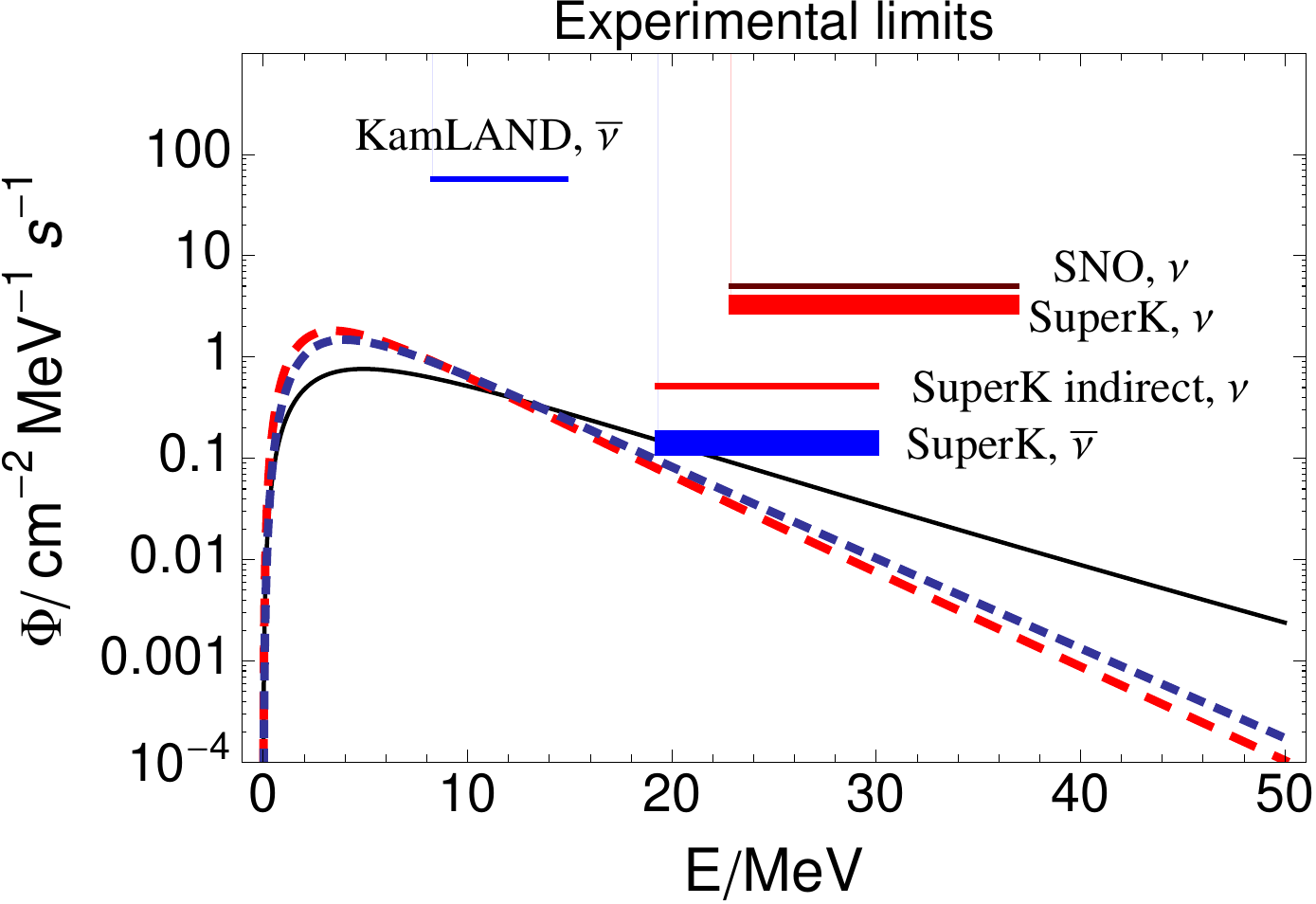}
\caption{Experimental limits for the $\nu_e$ and $\bar\nu_e$ components of the DSNB compared with theoretical predictions for three different models of neutrino spectra. See \cite{lunardini} for details.} 
\label{fig:lunardini}
\end{figure}

The DSNB has not been detected yet. Fig. \ref{fig:lunardini} compares several theoretical predictions with the experimental limits for $\nu_e$ and $\bar\nu_e$ components of the DSNB \cite{lunardini}. Limits on the energy-integrated fluxes have been divided by the size of the energy window of sensitivity of the experiment. At low energies, the best limit is from the KamLAND experiment \cite{kamland-dsnb} ($\Phi$($\bar\nu_e$) $<$  3.7 x 10$^2$ cm$^{-2}$s$^{-1}$ at 90\% CL for 8.3 $<$ E $<$ 14.8 MeV) and at high energies the strongest limit is on the $\bar\nu_e$ component from the SK experiment \cite{sk-dsnb} ($\Phi$($\bar\nu_e$) $<$ 1.2 cm$^{-2}$s$^{-1}$ at 90\% CL for E $>$ 19.3 MeV). The detection window is determined by the backgrounds. The SK limit is very close to the predictions and suggests the detectability of the DSNB at current or near future detectors.

In Water Cherenkov detectors, the main backgrounds are reactor neutrino fluxes and cosmic-ray induced spallation events at low energies and atmospheric neutrinos at higher energies. There is a window between 20-40 MeV where the DSNB dominates the neutrino flux incident on Earth. In the most recent work by SK \cite{sk2-dsnb}, the energy threshold has been lowered to 13.3 MeV using the tagging neutron capture on hydrogen and no signal was found. An intense R\&D is underway towards a gadolinium-enhanced SK providing a higher signal detection efficiency and greater background rejection to improve the SK sensitivity to the DSNB signal.

For LSc detectors, the main backgrounds are at low energies reactor neutrinos and at higher energies NC \& CC atmospheric neutrino interactions. In JUNO \cite{JUNO} for a fiducial mass of 17 kt a few events are expected per year. There is a window between 11 and 30 MeV where the DSNB dominates all backgrounds. The Pulse Shape Discrimination technique allows to remove fast neutrons and NC atmospheric neutrinos. JUNO may be able to detect the DSNB signal at 3$\sigma$ level after 10 years assuming a 5\% background uncertainty.

LAr TPCs would be able to mainly detect the $\nu_e$ component of the DSNB signal providing complementary information with respect to WC and LSc detectors. The main background sources for these events in the relevant neutrino energy range of 10-50 MeV are solar and low energy atmospheric neutrinos. Depending on the theoretical predictions for the DSNB flux, a 100 kt LAr detector running for five years would get more than 4$\sigma$ measurement of the DSNB flux \cite{cocco}.

\section{Summary}
The detection of SN neutrino events is one of the main goals of future large underground detectors. SN neutrinos can provide information about fundamental processes related to SN physics and neutrino properties. The understanding of the SN core-collapse mechanism and the neutrino flavor transformations in the star is still in progress. New experimental inputs are needed from the detection of SN neutrinos of all flavors. The main detector requirements to measure SN neutrinos are a good energy reconstruction and high trigger efficiency at low energies, good timing resolution, directionality and high event reconstruction efficiency. The complementarity between different detector technologies will be crucial to extract the maximal information from the next SN neutrino burst.


\end{document}